# FACTORS INFLUENCING THE EARTH'S MAGNETIC FIELD EVOLUTION

A. Yu. Kurazhkovskii[1], N. A. Kurazhkovskaya[2], and B. I. Klain[3]

*Borok Geophysical Observatory, Institute of the Physics of the Earth, Russian Academy of Sciences, Borok, Yaroslavl Region, 152742 Russia*

[1] E-mail:*ksasha@borok.adm.yar.ru*
[2] E-mail:*knady@borok. adm.yar.ru*
[3] E-mail:*klain@borok. adm.yar.ru*

**Abstract.** The relationship between the behavior of an ancient geomagnetic field characteristics (paleointensity and frequency of inversions) and cyclic recurrence of endogenic and cosmogeneous processes which are conceivably connected with radial mantle heat transmission and the Earth rotation speed has been studied. It is shown that endogenic processes affect the behavior of the paleointensity and frequency of inversions. Large basalt effusions identified with plumes are accompanied by the changes in paleointensity (by 30-40) % and the frequency of inversions. Characteristic time intervals of paleointensity variations caused by the formation of plumes makes up 10-20 Ma. The paleointensity varies (by 15-30) % according to phases of the riftogenesis activization and tension - compression cycles. The dependence of geomagnetic field behavior on changes of the Earth rotation speed which occurred as a result of the Earth - Moon - Sun system evolution has been analyzed. Thus, in accordance with phases of the Earth - Moon distance changes (periodicity about 200 Ma) average values of the paleointensity varied by a factor of two. Besides, the frequency of inversions, asymmetry of polarity and features of paleointensity variations changed too.

## 1. Introduction

According to modern concepts the generation of the planetary magnetic field requires the combination of three factors: the presence of the fluid core with good conductivity, radial flows of the fluid (thermal convection) and rotation of the planet round its axis. Some endogenic and cosmogeneous processes are apparently accompanied by changes in a radial heat transmission and the Earth rotation speed. The assumption that characteristics of an ancient magnetic field (paleointensity and reversal frequency) should vary in accordance with these endogenic and a cosmogeneous process seems to be true. However there is no consensus of opinion on the interdependence of endogenic and cosmogeneous processes and the behaviour of ancient magnetic field characteristics [*Dobretsov*, 1997; *Pechersky*, 2007]. Moreover even the conclusion about the existence of the interdependence of geomagnetic field characteristics [*Valet* 2003; *Tarduno et al*., 2006; *Kurazhkovskii et al*., 2008] has not become generally accepted.

The present publication deals with the results of studies on the relationship between the ancient geomagnetic field behavior and cycles of some endogenic and cosmogeneous processes.

## 2. Paleomagnetic data

The analysis of the paleointensity behavior was based on the results of the determination of the ancient magnetic field module from databases (DB) [*http//wwwbrk.adm.yar.ru/palmag/index.html*] and the author's data received on sedimentary rocks [*Kurazhkovskii et al*., 2008]. The DB contains results of the determination of the ancient magnetic field intensity obtained on thermomagnetized rocks. In many cases in addition to values of paleointensity (H) the DB includes values of the virtual dipole moment (V). Fig. 1 represents data V/Vo and H/Ho from the DB, where Vo = $8 \cdot 10^{22}$Am$^2$, Ho - value of the average modern magnetic field on the Earth surface (Ho=40 µT). As it is seen from Fig. 1 parameters V/Vo and H/Ho are given approximately identical picture of the geomagnetic field module behaviour. We shall analyze the paleointensity behaviour as the data on the virtual dipole moment values are less numerous.

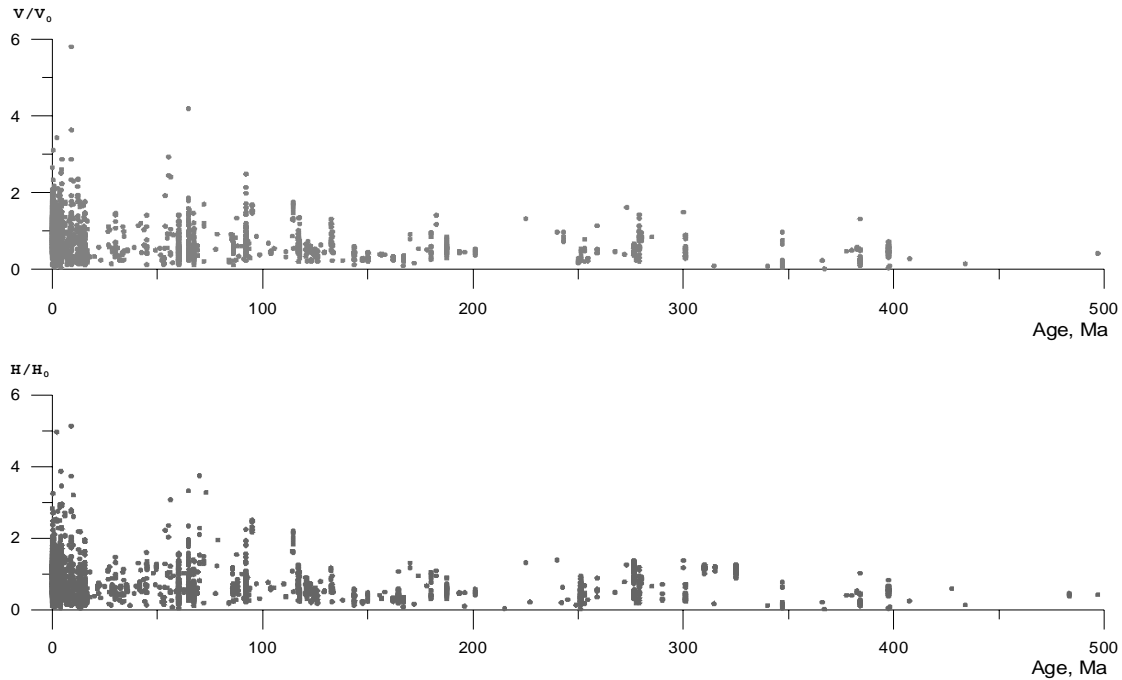

**Figure 1.** Values of the virtual dipole geomagnetic moment (V/Vo) and the ancient geomagnetic field (H/Ho) determined by termomagnetized rocks from the DB.

For determination of the inversions frequency behavior in the Phanerozoic we used the scale [*Supplements…*, 2000] which in many respects coincides with the scale [*Gradstein et al.*, 2004]. The frequency of inversions in an interval 120-85 Ma (Cretaceous Superchron mainly of the direct polarity) was estimated on the basis of the data [*Guzhikov et al.*, 2007]. It is necessary to note that there are some distinctions in views on behavior of the reversal frequency. So, according to *Gradstein et al.* [2004] there were no inversions in the interval 83-118 Ma. According to the data [*Supplements…*, 2000; *Guzhikov et al.*, 2007; *Molostovskii et al.*, 2007] inversions occurred, but they were few. Scales of polarity [*Supplements…*, 2000; *Gradstein et al.*, 2004] differ considerably in an estimation of the inversions frequency of some ages of the Jurassic period. In this connection a detailed comparison of the frequency of inversions and the paleointensity was made only for the last 160 Ma the scales of which [*Supplements…*, 2000; *Gradstein et al.*, 2004] differ insignificantly.

## 3. Behavior of the paleointensity and the frequency of inversions

The data received on relatively new [*Valet and Meynadier*, 1993; *Tauxe and Ymazaki*, 2007] and ancient [*Kurazhkovskii et al.*, 2008] sedimentary rocks testify that the amplitude and periodicity of the paleointensity variations changed chaotically. Characteristic time intervals of variations of the ancient geomagnetic field intensity which can be resolved by paleomagnetic methods range from tens of thousands (Fig. 2) up to tens and hundreds of millions years (Fig. 3).

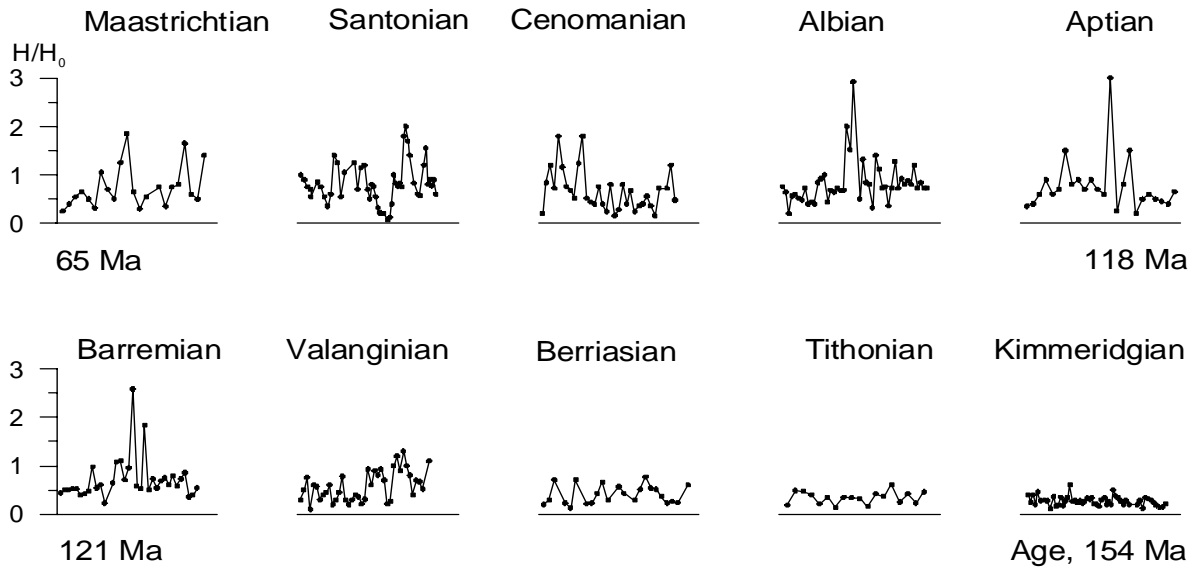

**Figure 2.** Fragments of behavior of the Jurassic - Cretaceous paleointensity obtained on sedimentary rocks.

The variations occuring with periodicity about tens of thousands years (Fig. 2) have the maximal amplitude (up to 3Ho). According to the DB the amplitude of paleointensity changes could reach 5Ho (Fig. 1). In works [*Valet*, 2003; *Tarduno et al.*, 2006; *Kurazhkovskii et al.*, 2008] it has been shown that frequency of inversions depends inversely on paleointensity and the amplitude of its variations. Changes of the inversions frequency and paleointensity for the last 170 Ma [*Kurazhkovskii et al.*, 2008] is presented in Fig. 3.

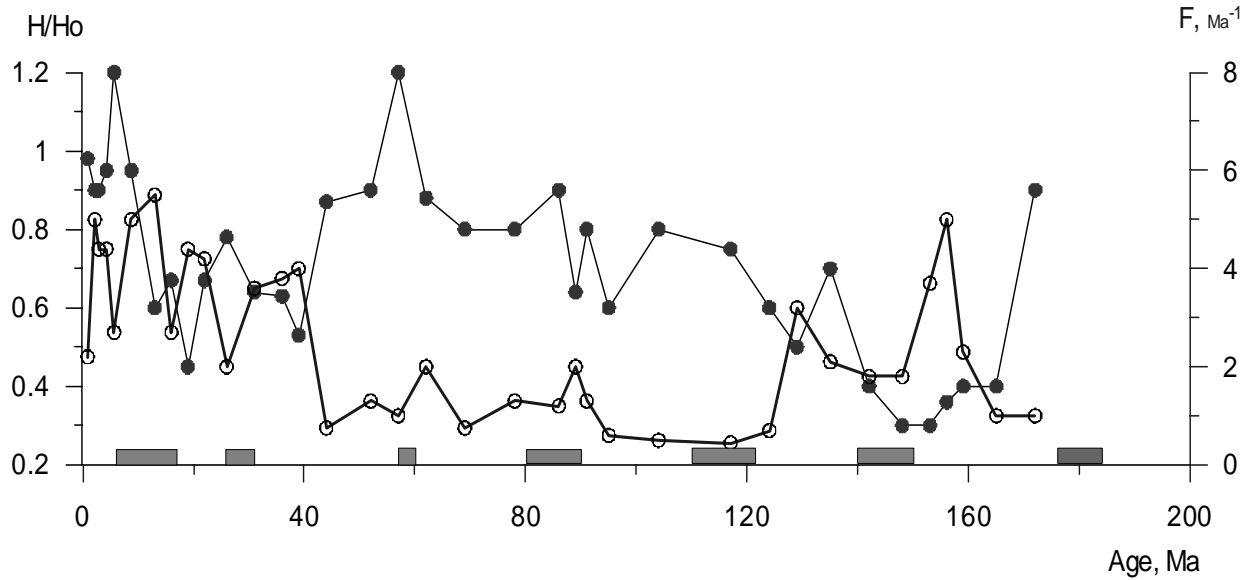

**Figure 3.** Changes in average for a geological age paleointensity values (H/Ho) (dark circles) and frequency of inversions (F) (light circles) in the last 170 Ma. Time intervals of the basalt effusions identified with plumes are shown by dark rectangles [*Dobretsov*, 1997; *Grachev*, 2000].

The same interrelation between characteristics of a geomagnetic field follows from the results of numerical modeling of *αω* - dynamo evolution [*Glatzmaier and Roberts*, 1995].

## 4. Conformity of behavior of the ancient geomagnetic field to some endogenic processes

Direct evidences of changes in thermal flows through a core – mantle border do not exist. However a nonuniform occurrence of geotectonic processes can serve as an indirect evidence of it. Intensification of the riftogenesis and large lava flows on the Earth's surface, probably, should be preceded (or accompanied) by intensification of heat transmission to the Earth's surface. Thus large effusions of basalts and riftogenesis can be considered as factors capable to influence the behavior of the geomagnetic field characteristics.

Data on significant effusions of basalt lavas on the Earth's surface with which the plumes are identified [*Dobretsov*, 1997; *Grachev*, 2000] are presented in Fig. 3. As it is seen from the figure these effusions are accompanied by approximately identical changes in behavior of the geomagnetic field. At first paleointensity decreases (in the beginning or before appearance of the plume on the Earth's surface) and then increases on average by (0.3 - 0.4) Ho. The frequency of inversions increases before or in the beginning of the magmatic event and decreases at the end. The characteristic time intervals of changes in paleointensity and the frequency of inversions connected with these endogenic processes make up approximately 10 – 20 Ma. If effusions of magma occur frequently (in the interval of 110 - 130 Ma three such events have been marked [*Dobretsov*, 1997; *Grachev*, 2000]) relatively short cycles (about 10 Ma) of the paleointensity changes merge into one longer cycle. At increasing and high values of paleointensity in the Cretaceous- Paleogene basalt effusions occur rather frequently (approximately in 15 Ma). At decreasing and low values of paleointensity (in the Eocene and Jurassic) the time interval between events of plumes increases up to 30 Ma. Thus, intensification of magmatic activity is accompanied by an increase of paleointensity.

It should be noted that the data on paleointensity and frequency of inversions have been obtained on the basis of different methods and geological objects. At the same time the both characteristics of the geomagnetic field change consistently their behavior before large basalt effusions. Thus, two different sources of the data on behavior of the geomagnetic field testify to the existence of the relation between endogenic and geomagnetic processes.

The number of paleomagnetic data decreases sharply with deepening into geological history. Nevertheless it is possible to make an attempt to consider the relationship between the most significant geological events and behaviour of the geomagnetic field in Paleozoic and Precambrian using very limited data. One of the most significant magmatic events of the Phanerozoic is the formation of the Siberian traps (250 Ma). The available data are not sufficient for conclusions about the dynamics of the paleointensity (Fig. 1), but the behaviour of frequency of inversions can be analyzed using the scales [*Supplements…*, 2000; *Gradstein et al.*, 2004]. The same way as in the Cretaceous – Cenozoic periods the frequency of inversions had increased (265 Ma ago) before the formation of traps and decreased after their effusions (245 Ma). According to numerous data, for example, [*Sorokhtin*, 2007] the most intensive magmatic processes in the Earth's history occurred 2.6 Ga. The data on the Precambrian paleointensity [*Smirnov et al.*, 2003; *Macouin and Valet*, 2004; *Tauxe and Ymazaki*, 2007] testify to a considerable increase of the paleointensity close to this period. Besides higher values of the paleointensity observed 1.7 and 1.4 - 0.8 Ga [*Tauxe and Ymazaki*, 2007] can be connected with an increase in tectonic activity happened on the verge of the Proterozoic epoch.

The comparison of paleointensity behaviour with cyclic recurrence of riftogenesis phases presents difficulties because of the absence of general agreement concerning their time boundaries and even the existence of cyclic recurrence. The monograph by *Khain and Lomize* [2005] presents one of the variants of fragmentation of the geological history for the last 200 Ma according to riftogenesis activity. The duration of phases of the riftogenesis activity is tens millions years. For such long time intervals the paleointensity varies in a complicated way and within rather wide ranges (Fig. 2, 3). In all cases the paleointensity increased by 15-30 % during phases of high activity of riftogenesis [*Kurazhkovskii et al.*, 2007].

In the work by *Milanovsky* [1996] the geological history is subdivided into short (on average less than 10 Ma) intervals characterized by a primary stretching or compression of the Earth's crust. We have determined average values of the paleointensity corresponding to intervals either of primary stretching or compression for the last 170 Ma (20 intervals) using the DB. Apparently, the accuracy of determination of a volcanogenic age used for the paleointensity determination in some cases becomes commensurable with duration of the intervals. Nevertheless the paleointensity corresponding to the intervals of a primary stretching on average was higher by 20 % than the paleointensity corresponding to the intervals of compression.

Thus the changes in characteristics of the geomagnetic field are interdependent on endogenic processes the results of which are large basalt effusions and cycles of riftogenesis.

## 5. The relation of the geomagnetic field behavior with cosmogeneous factors

Changes in characteristics of the Earth orbital movement and evolution of the Earth - Moon - Sun system are usually considered as cosmogeneous factors capable to influence the geomagnetic field behavior.

The comparison of the paleointensity behavior with orbital cycles (known as the Milankovich's cycles) was carried out repeatedly that is mentioned in the review by *Valet* [2003]. From results of these works it follows that the paleointensity behaviour of the Late Cenozoic is characterized by the spectrum of variations with typical periods close to orbital cycles. At the same time between variations of the paleointensity and orbital cycles there is no exact phase conformity [*Valet*, 2003]. The paleointensity variations in the Late Cenozoic [*Valet and Meynadier*, 1993], Cretaceous and Late Jurassic (Fig. 2) differ considerably by the amplitude and the structure. Changes of the geomagnetic field intensity with characteristic time intervals of tens of thousands years cannot be explained by the influence of only cosmogeneous factors. Similar cyclic recurrence [*Seliverstov*, 2001] is also peculiar to endogenic processes. In the work [*Kurazhkovskii et al.*, 2007] it is shown that the correlation exists between of the paleointensity variations and volcanism activity. Besides, the occurrence of paleointensity bursts which are observed from Barremian to Maastrichtian (Fig. 2) testifies to the fact that similar characteristic time intervals are peculiar to turbulent processes arising in a fluid core. Hence the existence of the direct relationship between the paleointensity behavior and orbital cycles is improbable.

The change of the distance between the Earth and the Moon by far influences the speed of the planet rotation and should influence the behavior of the geomagnetic field characteristics. Cyclic recurrence of the Earth – Moon distance changes can be calculated with good accuracy for remotest time intervals [*Avsyuk*, 1993]. Values of the paleointensity and the frequency of inversions calculated according to phases (approach – moving off) of the changes of the Earth - Moon distance are shown in Fig. 4.

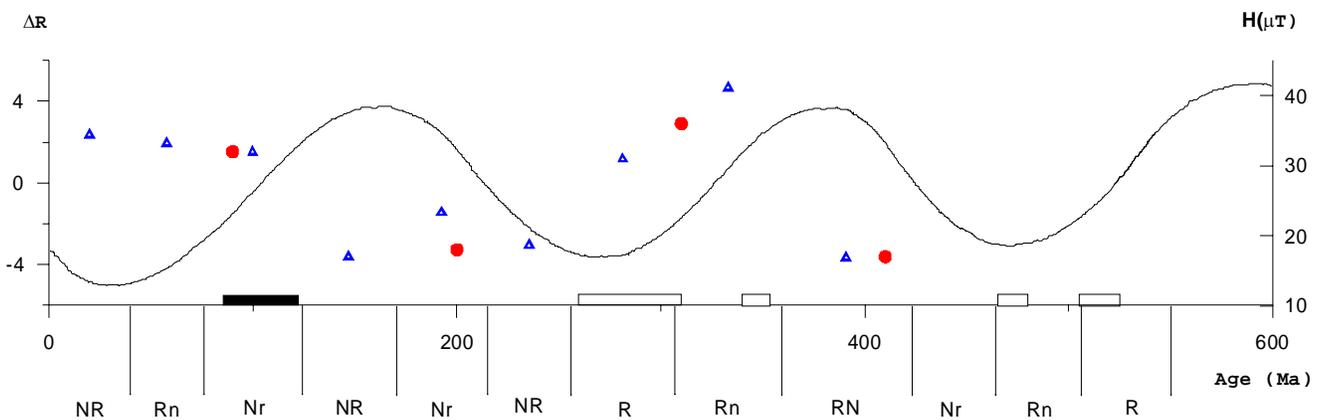

**Figure 4.** The relation of the paleointensity changes with changes of the Earth - Moon distance (ΔR). Points indicate average values of the paleointensity from the DB for a phase of the change of the Earth - Moon distance (approach – moving off) [*Avsyuk*, 1993]. Triangles indicate average paleointensity values corresponding to polarity Superchrones. Time boundaries of the polarity Superchrones [*Supplements…*, 2000] are shown on the abscissa axis by vertical lines. Superchrones mainly with one polarity are marked by rectangles.

At closer approach of the Earth and the Moon the paleointensity and the asymmetry of polarity increased, and the frequency of inversions decreased. Superchrones with obvious prevalence of one polarity were observed only in phases of the approach of the Earth and the Moon [*Supplements…*, 2000; *Gradstein et al.*, 2004].

The increase of average values and amplitudes of paleointensity variations was observed during the Carboniferous - Permian and Cretaceous - Paleogene intervals. Average values of the paleointensity varied by a factor of two depending on a phase in the cycles of the Earth - Moon distance.

On the basis of the found relation between the paleointensity behavior, frequency of inversions and changes of the Earth - Moon distance it is possible to estimate probable average values of the Ordovician - Cambrian paleointensity at a qualitative level. The combination of such factors, as: approach of the Earth and Moon, existence of Superchrones mainly of one polarity and relatively low frequency of inversions [*Supplements…*, 2000; *Gradstein et al*., 2004] allows us to assume that average values of the paleointensity in this interval should be relatively high.

## 6. Conclusions

Changes of the geomagnetic field characteristics occurring with typical time intervals of approximately millions and tens of millions years can be possibly explained by the effect of cosmogeneous and endogenic factors. So, according to phases of changes in the Earth - Moon distance (periodicity about 200 Ma) average values of the paleointensity vary by a factor of two and the frequency of inversions, asymmetry of polarity and features of paleointensity variations also change. Endogenic processes influence the behaviour of the paleointensity and frequency of inversions too. Large effusions of basalts identified with plumes are accompanied by changes of the paleointensity by (30-40) %. Characteristic time intervals of the paleointensity variations related to the formation of plumes amount to 10-20 Ma. The paleointensity varies by (15-30) % according to phases of the riftogenesis activation and stretching - compression cycles.

Variations of the paleointensity with characteristic time intervals up to tens of thousands years have large amplitudes (reach 5Ho), but they cannot be connected only with the effect of cosmogeneous, or endogenic factors. Similar periodicity is found in changes of the Earth orbit elements, in manifestation of the volcanic activity and probably takes place in the processes of turbulence in a fluid core.